\documentclass[pdflatex,sn-mathphys-num,twocolumn]{sn-jnl}
\usepackage{graphicx}%
\usepackage{multirow}%
\usepackage{amsmath,amssymb,amsfonts}%
\usepackage{amsthm}%
\usepackage{mathrsfs}%
\usepackage[title]{appendix}%
\usepackage{xcolor}%
\usepackage{textcomp}%
\usepackage{manyfoot}%
\usepackage{booktabs}%
\usepackage{algorithm}%
\usepackage{algorithmicx}%
\usepackage{algpseudocode}%
\usepackage{listings}%
\usepackage{geometry}
\usepackage{lineno}
\begin{document}
\title[Three-dimensional sizes and shapes of pion emission in heavy-ion collisions]{Three-dimensional sizes and shapes of pion emission in heavy-ion collisions}

\author*[1]{\fnm{Dániel} \sur{Kincses}}\email{kincses@ttk.elte.hu}
\author[1]{\fnm{Emese} \sur{Árpási}}
\author[1]{\fnm{László} \sur{Kovács}}
\author[1]{\fnm{Márton} \sur{Nagy}}
\author[1]{\fnm{Máté} \sur{Csanád}}

\affil[1]{\orgdiv{Department of Atomic Physics}, \orgname{ELTE Eötvös Loránd University}, \orgaddress{\street{Pázmány Péter sétány 1/A}, \city{Budapest}, \postcode{H-1117}, \country{Hungary}}}

\abstract{In the era of precision measurements in high-energy heavy-ion physics, there is an increasing expectation towards phenomenological and theoretical studies to provide a better description of data. In recent years, multiple experiments have confirmed through two-pion Bose-Einstein correlation measurements that the shape of the two-pion pair source can be well described by Lévy-stable distributions. However, direct comparisons of new phenomenological results with the data are still needed to understand the underlying phenomena and learn more about the nature of pion emission. In this paper, we present a three-dimensional analysis of the two-pion source in Monte-Carlo simulations of $\textnormal{Au}{+}\textnormal{Au}$ collisions at 200 GeV per nucleon collision energy based on EPOS3, and discuss a detailed comparison with the most recent centrality-dependent measurements from the PHENIX Collaboration. We find that the model remarkably agrees with the data for peripheral collisions, while an increasing discrepancy is found in case of central collisions for the source shapes and sizes. On the other hand, scaled correlation strengths are found to be well described by the model across all investigated system sizes.}



\maketitle

\section{Introduction}\label{s:intro}

The study of two-pion relative momentum correlations has been fundamental to heavy-ion physics since Goldhaber et al.\@ first observed that pion pairs show an enhancement at low relative momenta~\cite{Goldhaber:1960sf} - a consequence of the Bose--Einstein symmetrization of identical bosons. Since then, a plethora of results have been published, and a whole sub-field of high-energy physics research has arisen called femtoscopy~\cite{Lisa:2005dd,Lednicky:2005af,Csanad:2024hva}. The premise of this field is that relative momentum correlations are connected to spatial correlations (also called the pair source function) through the pair wave function that contains quantum statistical and final-state interaction effects. This connection can be utilized to reconstruct the pair source in heavy-ion reactions and to learn about its space-time geometry. 

As the experimental data became more precise, it was observed that the pion pair source exhibits features of a long tail~\cite{PHENIX:2007grx,Brown:2000aj}. A simple Gaussian distribution could not provide an accurate and statistically acceptable description of two-pion correlations, and thus many experiments started to utilize L\'evy-stable distributions as a source shape assumption~\cite{PHENIX:2024vjp,Kincses:2024sin,NA61SHINE:2023qzr,CMS:2023xyd}. In recent years, many phenomenological investigations have also been published~\cite{Kincses:2024lnv,Csanad:2024jpy,Nagy:2023zbg,Korodi:2022ohn,Kincses:2019rug,Kincses:2025iaf}, related to the L\'evy-stable shape of the pion source. In Ref.~\cite{Kincses:2024lnv}, it has been shown that in Au+Au collisions at $\sqrt{s_{\rm NN}}=200\textnormal{ GeV}$, the main phenomenon behind the apparent long tail is the so-called L\'evy-walk of hadrons during the hadronic rescattering phase of the collisions. This means that in the expanding hadron gas, elastic and inelastic collisions and decays of resonances contribute to the appearance of long steps in pion paths and thus the long tail in the pair distance distribution. Using EPOS~\cite{Werner:2010aa}, a state-of-the-art Monte-Carlo-based simulation model containing both a hydrodynamic evolution and a hadronic rescattering phase, in Ref.~\cite{Kincses:2024lnv}, the shape of the pion pair source has been directly investigated. It has been found that a three-dimensional elliptically contoured L\'evy-stable distribution can provide a good description of the reconstructed pion pair source on an event-by-event basis. 

In this follow-up study, we continue along the lines of Ref.~\cite{Kincses:2024lnv}, and present a detailed comparison of the centrality dependence of the pion pair source parameters from EPOS, and from the recently published PHENIX measurements~\cite{PHENIX:2024vjp}. The structure of the paper is the following. In Section~\ref{s:methods} we discuss the details of the simulation and the methods of the pair source reconstruction and extraction of the source parameters. In Section~\ref{s:results}, we present a detailed comparison of the simulation results with the PHENIX measurements and discuss the consequences of our findings. Finally, in Section~\ref{s:summary} we summarize and conclude.

\section{Methods}\label{s:methods}

For this study, a sample of 300,000 minimum bias Au+Au events were simulated at the center-of-mass collision energy of $\sqrt{s_{\rm NN}} = 200\textnormal{ GeV}$ with the EPOS model (version 359). Details about the model can be found in Refs.~\cite{Werner:2007bf,Werner:2010aa,Werner:2013tya,Kincses:2024lnv}. Using the freeze-out coordinates of identically charged pion pairs created in the simulated events, we calculated the three-dimensional spatial separation vector $\boldsymbol{\rho}$ boosted to the Longitudinal Co-Moving System (LCMS)~\cite{Csorgo:1991ej}. As described by Equations 13-15 of Ref.~\cite{Kincses:2024lnv}, the components of $\boldsymbol{\rho}$ were calculated in the out-side-longitudinal coordinate system~\cite{Grassberger:1976au,Podgoretsky:1982xu,Bertsch:1993nx,Pratt:1995wm}, where the 'out' direction is along the average transverse momentum of the pair, 'long' is the beam direction and 'side' is perpendicular to the other two. The single particle selection criteria applied for transverse momentum was ${0.15 < p_T\;(\textnormal{GeV}/c) < 1.0}$, and for pseudorapidity $|\eta| < 1$. To use only pairs relevant for experimental momentum correlation measurements, we applied a two-track selection criterion for the magnitude of the relative momentum variable ${|\boldsymbol{q_{\rm LCMS}}|=Q_{\rm LCMS}}$ (defined in Equations~26-28 of Ref.~\cite{PHENIX:2017ino}). The choice for this selection was the following: ${Q_{\rm LCMS} < \sqrt{(0.15\textnormal{ GeV})\;m_T}}$, where $m_T=\sqrt{m_\pi^2+k_T^2}$ is the average transverse mass of the pair, and $k_T$ is the average transverse momentum of the pair (the projection of the average momentum perpendicular to the beam direction). This relative momentum selection approximately corresponds to the usual fit limits applied in experimental analyses. 

Following along the lines of Ref.~\cite{Kincses:2024lnv} we constructed one-dimensional projections of the three-dimensional $D(\boldsymbol{\rho})$ pair source distribution along the out-side-longitudinal axes, in 11 different $50\textnormal{ MeV}/c$ wide $k_T$ classes, ranging from $0.175\textnormal{ GeV}/c$ up to $0.725\textnormal{ GeV}/c$. To investigate centrality dependence, we used six 10\% wide centrality classes, between $0-60\%$.

\subsection{The core-halo picture}

In experiments, the above mentioned spatial pair source distributions cannot be measured directly (as spatial resolutions limit the observation of sizes already well above the femtometer scale). The method of femtoscopy still allows to infer the spatial geometry of the source experimentally, via the~\cite{Goldhaber:1960sf,Kopylov:1974th,Podgoretsky:1989bp,Koonin:1977fh,Yano:1978gk,Lednicky:1981su,Pratt:1984su,Pratt:1997pw,Lisa:2005dd}
\begin{align}
 C(\boldsymbol{q}) = \int D(\boldsymbol{\rho}) |\Psi_{\boldsymbol{q}}(\boldsymbol{\rho})|^2 d^3\boldsymbol{\rho}
 \label{e:cq}
\end{align}
relation, where $\Psi_{\boldsymbol{q}}(\boldsymbol{\rho})$ is the pair-wavefunction at relative momentum $\boldsymbol{q}$ and separation $\boldsymbol{\rho}$. Let us note that the variables of the pair-wave function have to be taken in the pair rest frame, or pair co-moving system (PCMS); the connection between LCMS and PCMS is detailed, for example, in Ref.~\cite{Kurgyis:2020vbz}. If the final-state strong and electromagnetic (Coulomb) interactions are neglected (denoted by a $(0)$ superscript), then ${\Psi_{\boldsymbol{q}}^{(0)}(\boldsymbol{\rho})=1+\cos(\boldsymbol{q}\boldsymbol{\rho})}$, thus at zero relative momentum ($\boldsymbol{q}=0$) the value of the correlation function is by definition equal to two: ${C_2^{(0)}(\boldsymbol{q}=0) = 2}$. This is an approximation valid for thermal, completely chaotic particle emission. Experimentally, however, the $\boldsymbol{q}=0$ limit cannot typically be reached, since the two-particle momentum resolution sets a lower bound on the accessible $\boldsymbol{q}$ region. A minimal accessible value $q_{\rm min}$ is typically around 5--10~MeV/$c$, and depends on the average momentum of the particles. This corresponds to a spatial resolution of $\hbar / q_{\rm min} \approx 25\text{--}30$ fm. Spatial structures larger than this contribute to a region in the correlation function where it is not well measurable. This in turn leads us to the definition of the correlation strength (or correlation intercept) parameter $\lambda$:
\begin{align}
 \lambda = C_2^{(0)}(\boldsymbol{q} \rightarrow 0) - 1,
\end{align}
where the arrow represents the extrapolation, based on the experimentally accessible $\boldsymbol{q}$ region. Note that for charged particles, $\Psi_{\boldsymbol{q}}$ has to be utilized instead of $\Psi_{\boldsymbol{q}}^{(0)}$. In this case, $\lambda$ can be experimentally defined through the so-called Coulomb-corrected correlation function, as discussed e.g. in Ref.~\cite{PHENIX:2017ino}, and it can be obtained as a parameter of the fit function derived from Eq.~\eqref{e:cq}.

The correlation intercept parameter can be understood within the core-halo picture~\cite{Alt:1999cs,Csorgo:1994in,PHENIX:2017ino}, where the source is composed of two parts. These are
\begin{itemize}
\item the {\bf core}, composed of primordial pions, as well as those stemming from decays of short-lived resonances (such as $\rho$, $\Delta$, $K^*$, $\omega$, $\phi$), and
\item the {\bf halo}, composed of pions coming from decays of longer-lived resonances (such as $\eta'$, $\eta$, $D$, $K_0^{\rm S}$, $\Sigma$, $\Lambda$).
\end{itemize}
In this case, the spatial probability density of creating a particle at a given location can be decomposed as ${S=S_{\rm core} + S_{\rm halo}}$ (with $\int S = 1$). Henceforth, the pair source distribution $D$ (normalized as $\int D = 1$), which arises naturally as the autoconvolution of $S$, decomposes as
\begin{align}
 D = D_{\rm (c,c)} + D_{\rm (c,h)} + D_{\rm (h,h)},
\end{align}
where the terms on the right-hand side of the equation denote the contribution of core-core, core-halo and halo-halo pairs, respectively (note that the integral of any of the terms on the right-hand side of the above equation are less then unity; unlike the convention adopted in many previous studies, notably Ref.~\cite{PHENIX:2017ino}, where these terms are also normalized to unity). If the halo pions indeed come from a spatially distant region, then their momentum-space version (transformed via $\Psi_{\boldsymbol{q}}$) is experimentally unresolvable, and the correlation function in the accessible $\boldsymbol{q}$ region becomes~\cite{PHENIX:2017ino,Kincses:2019rug}
\begin{align}
C(\boldsymbol{q}) = 1 - \lambda + \int D_{\rm (c,c)}(\boldsymbol{\rho}) |\Psi_{\boldsymbol{q}}(\boldsymbol{\rho})|^2 d^3\boldsymbol{\rho}
\end{align}
where $\lambda = \int D_{\rm (c,c)}$ (note again that often a different normalization is used, where the integral of $D_{\rm (c,c)}$ is unity, as for example in Eq. (44) of Ref.~\cite{PHENIX:2017ino}), and thus $1{-}\lambda = \int \left(D_{\rm (c,h)} + D_{\rm (h,h)}\right)$. In the absence of final-state interactions, $|\Psi_{\boldsymbol{q}}(\boldsymbol{\rho})|^2 = 1 + \cos(\boldsymbol{q}\boldsymbol{\rho})$, thus one obtains
\begin{align}
C^{(0)}(\boldsymbol{q}) = 1 - \lambda + \int D_{\rm (c,c)}(\boldsymbol{\rho}) \left[1+\cos(\boldsymbol{q}\boldsymbol{\rho})\right] d^3\boldsymbol{\rho}
\end{align}
and thus extrapolating to zero in the experimentally accessible momentum range gives
\begin{align}
C^{(0)}(\boldsymbol{q} = 0) = 1 - \lambda + 2\int D_{\rm (c,c)}(\boldsymbol{\rho}) d^3\boldsymbol{\rho} = 1 + \lambda,
\end{align}
recovering the limiting behaviour mentioned above. This confirms that the experimental intercept parameter $\lambda$ can indeed be interpreted as the integral of $D_{\rm (c,c)}$, the core-core contribution of the pair source. In the subsequent sections, we thus normalize the obtained pair distributions to unity, fit the core-core part with a functional form, and obtain $\lambda$ as the integral of the fitted distribution.

\begin{figure}
    \centering
    \includegraphics[width=\linewidth]{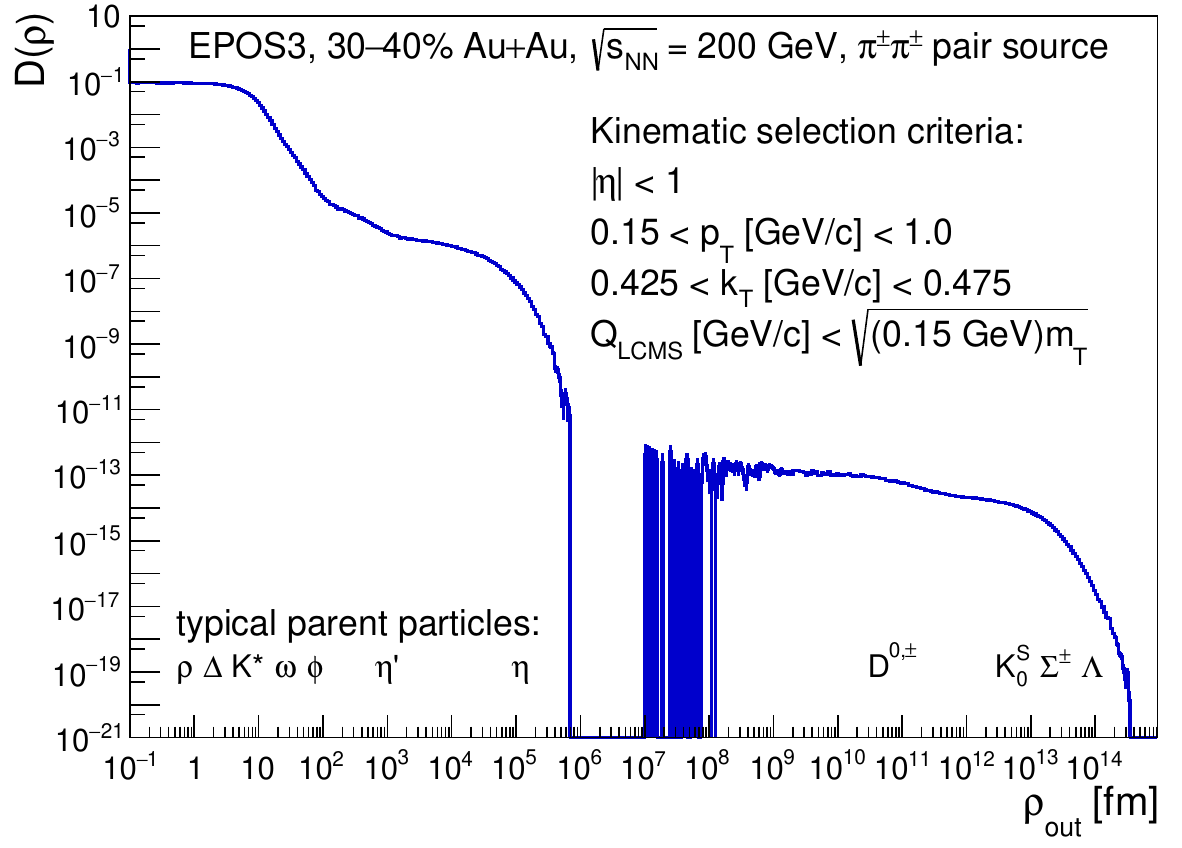}
    \caption{An example $D(\boldsymbol{\rho})$ distribution for 30--40\% centrality and $k_T$ within $[0.425,0.475]$ GeV/$c$, projected in the \emph{out} direction. The kinematical track ($\eta$, $p_T$) and pair ($k_T$, $Q_{\rm LCMS}$) selection criteria are also shown. Above the horizontal axes typical parent particles---producing the vast, $>99\%$ majority of pions---in the given $\rho$ range are indicated.}
    \label{f:examplerho}
\end{figure}

An example $D(\boldsymbol{\rho})$ distribution for 30--40\% centrality (approx. thirty thousand events) and $k_T$ within $[0.425,0.475]$ GeV/$c$, projected in the \emph{out} direction is shown in Fig.~\ref{f:examplerho}. Here it is apparent that the above-mentioned core-halo distinction exists: long-lived resonances contribute to great distances, and the central, ``core'' part is contributed to by shorter-lived resonances. It is also apparent that the part below $~100$ fm distance can be described by a single distribution, as discussed in the next subsection. Note furthermore that there is a ``hole'' around $10^6-10^7$ fm (1-10 nm) at the present level of statistics, this is due to the gap between the lifetime of the $\eta$ and the $D$ mesons, i.e., the lack of decay pions from either particle at the given number of events. We obtained similar $D(\boldsymbol{\rho})$ distributions in all of the six centrality classes from 0\% to 60\% (where we utilized the internal EPOS3 centrality definition) and 11 bins of $k_T$ from 0.175 to 0.725 GeV/$c$, for all three projections (out, side, long).

\subsection{Fitting method}

\begin{figure*}
    \centering
    \includegraphics[width=\linewidth]{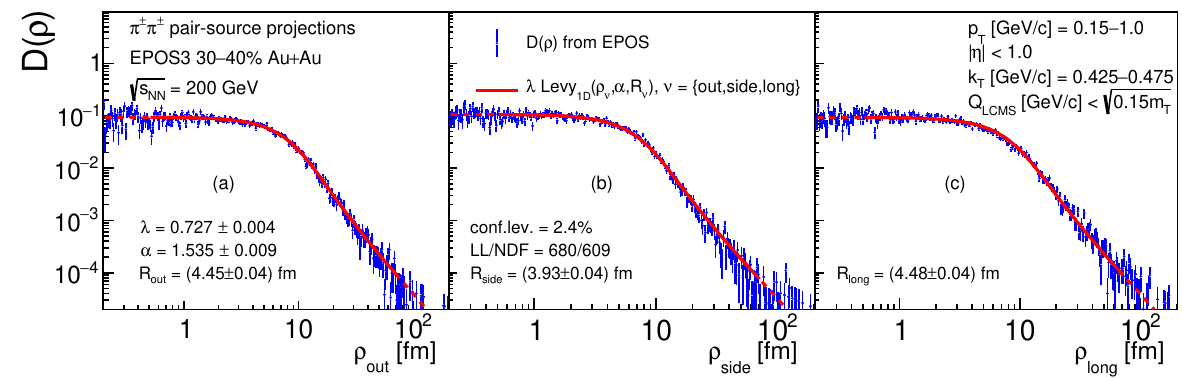}\\
    \includegraphics[width=\linewidth]{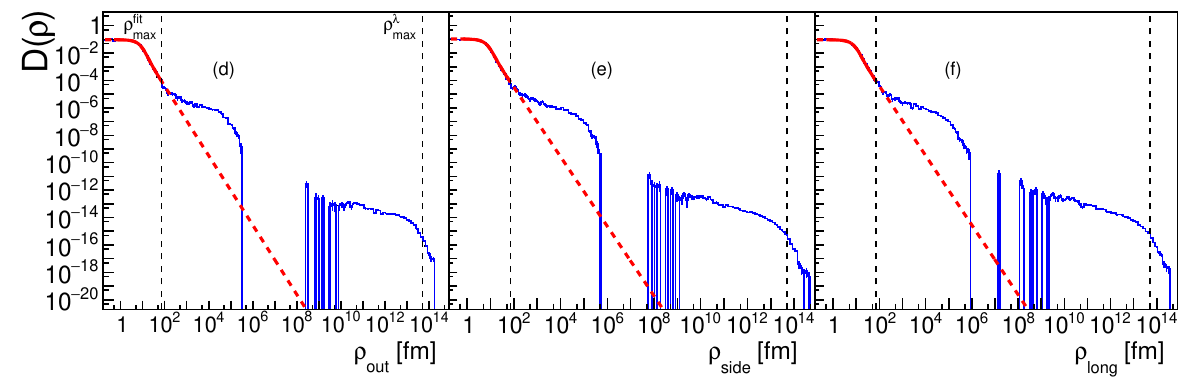}
    \caption{An example fit to projected $D(\boldsymbol{\rho})$ distributions for 30--40\% centrality and $k_T$ within $[0.425,0.475]$ GeV/$c$.  The solid red line corresponds to the fit, while the dashed version is its extrapolation. The blue histogram shows the EPOS3 $D(\boldsymbol{\rho})$ distribution. The range of the horizontal axis in panels (a), (b), and (c) (in the out, side, long directions, respectively) is slightly larger than the fitted region, while in panels (d), (e), and (f) the entire investigated region (up to a meter distance) is shown (in the out, side, long directions, respectively). The kinematical selection criteria, the fit parameters, their uncertainties, as well as the fit quality are indicated on the plots as well.}
    \label{f:examplefit}
\end{figure*}

We apply the same method described in Refs.~\cite{Kincses:2024lnv,Kincses:2025iaf}, fitting one-dimensional L\'evy-stable distributions simultaneously to the three projections in each $k_T$ class, then extracted the $k_T$ dependence of the fitted source parameters. The fit function was defined as:
\begin{align}
    &D_{\rm (c,c)}(\rho_\nu) = \lambda\;\mathcal{L}^{1D}(\rho_\nu, \alpha, R_\nu), \textnormal{ where}\label{e:Dccdef}\\
    &\mathcal{L}^{1D}(\rho_\nu, \alpha, R_\nu) = \frac{1}{2\pi}\int d\xi\, e^{i\xi\rho_\nu} e^{-|\xi R_\nu|^{\alpha}},
    &\rho_\nu = \boldsymbol{e}^{(\nu)}\cdot\boldsymbol{\rho},\\ 
    &R_\nu = \sqrt{\boldsymbol{e}^{(\nu),T}R^2\boldsymbol{e}^{(\nu)}},\\
    &R^2 = \textnormal{diag}\left(R_{\rm out}^2,R_{\rm side}^2,R_{\rm long}^2\right)\quad\textnormal{and}\\
    &\boldsymbol{e}^{(\nu)}\textnormal{ is a unit vector with }\nu = \rm{out, side, long}.\nonumber
    \label{e:1dlevy}
\end{align}
The two parameters characterizing such a distribution are called the L\'evy-exponent $\alpha$ describing the asymptotical behavior~\cite{Nolan:Levy} (the steepness of the power-law tail), and the L\'evy-scale $R$. The fit utilized log-likelihood minimization, preferred over conventional $\chi^2$ due to the limited number of entries in the majority of the bins considered for the fit.

In previous analyses~\cite{Kincses:2022eqq,Korodi:2022ohn,Kincses:2024lnv}, single-event spatial pair distributions were fitted with the above distribution (or its angle-averaged counterpart), and then the fit parameters from single-event fits were averaged. However, in the case of peripheral collisions or large pair transverse mass bins, a single event may not contain enough pairs to perform such single-event fits. We furthermore found that depending on event activity, the averaged fit parameters may change if $D(\boldsymbol{\rho})$ distributions from more events are summed before fitting. We determined a limit of convergence in terms of the $N_{\rm events}$ (number of events) summed, where the fit parameters do not change anymore, even if orders of magnitude more events are summed. This limit varies from a few events to several hundred, depending on centrality and pair transverse mass. Subsequently, we used the obtained $N_{\rm events}$ value, and included its variations in the systematic uncertainties.

An example fit for one centrality and pair transverse mass bin is shown in Figure~\ref{f:examplefit}, zoomed to the fit region in panels (a)-(c), and for the entire investigated region in panels (d)-(f). In this particular case, the optimized $N_{\rm events}$ value was 300. The fit function is $D_{\rm (c,c)}$ as defined Eq.~\eqref{e:Dccdef}, the core-core part of the pair source---its integral is thus $\lambda$, and the corresponding interaction-free momentum-space correlation function, $C^{(0)}(\boldsymbol{q})$, would in the experimentally accessible range tend to $1+\lambda$, as discussed above. It is readily apparent that the normalized total pair source (with $\int D = 1$) is above the fit function, especially in the tail region (above distances $~100$ fm), resulting in a $\lambda$ value smaller than unity. It is important to note furthermore, that in such fits there are two important range limits: (1) $\rho^{\rm fit}_{\rm max}$ corresponds to the upper limit of the fitted region, (2) $\rho^{\lambda}_{\rm max}$ corresponds to the range where the integral of $D$ is normalized to achieve $\int D = 1$. One could set $\rho^{\lambda}_{\rm max} = \infty$, but in an experimentally realized scenario, pions created (usually from decays) outside the detector volume are discarded by tracking. Thus if $D$ is to correspond to pairs made up of all pions considered in an experimental setting, we have to limit the normalization range in $\rho$. In this case, $\rho^{\lambda}_{\rm max}$ corresponds to this upper normalization limit. In an experimental comparison of the calculations, a $\rho^{\lambda}_{\rm max}$ shall be chosen that corresponds to the track resolution, or for example, a largest possible ``Distant of Closest Approach'' (DCA) that is allowed in the experimental pion sample. Essentially, the integration limit $\rho^{\lambda}_{\rm max}$ at these large distances matters---perhaps contrary to expectations---because there are particles decaying to pions at these distances, producing far-away pion pairs. Depending on the experimental apparatus, these pions are discarded from the investigated sample or not even registered by tracking (for example, if the decay producing one of the pions happens at a distance of dozens of centimeters from the vertex), or they are included (if the decay happens at a few mm or cm distance, or if the experiment does not apply a strict DCA selection).

\subsection{Systematic uncertainties}
When considering the above fits, a number of choices have been made regarding analysis settings. These are the number of events averaged ($N_{\rm events}$), the pair kinematic limit ($Q_{\rm LCMS}^{\rm max}$), the $\rho$ fit range, and the normalization (integration) limit ($\rho^\lambda_{\rm max}$). The default (strict, loose) settings are given in the below list:
\begin{itemize}
\item $N_{\rm events}$: 
  \begin{itemize}
  \item 0-10\%: 50 (10, 30,000)
  \item 10-20\%: 100 (75, 30,000)
  \item 20-30\%: 200 (150, 30,000)
  \item 30-40\%: 300 (200, 30,000)
  \item 40-50\%: 1000 (600, 30,000)
  \item 50-60\%: 3000 (1000, 30,000)
  \end{itemize}
\item $Q_{\rm LCMS}^{\rm max}$: $\sqrt{A\cdot m_T}$ with $A =$ 0.15 (0.05, 0.25) [GeV/$c$]
\item $\rho^{\rm fit}_{\rm max}$: $\sqrt{B/m_T}$, $B =$ 2500 (1600, 3600) [fm$^2\cdot$GeV/$c^2$]
\item $\rho^\lambda_{\rm max}$: this only affects $\lambda$, and we consider it a genuine physical parameter, not a ``nuisance parameter''. Thus, we show results with various choices for this setting, ranging from 1 mm to 50 cm, corresponding to typical experimental tracking capabilities (where tracks created far from the main interaction vertex are not reconstructed or discarded).
\end{itemize}
The total systematic uncertainty is then the squared sum of the uncertainties of the above sources (with the exception of $\rho^\lambda_{\rm max}$, for which we quote all results separately below). We furthermore considered the centrality definition: besides the default setting in EPOS3, we also determined centrality limits based on the number of charged hadrons, as customary in experimental analysis. This resulted in a negligible ($<1\%$) change in the parameters.

Note that, unlike Refs.~\cite{Kincses:2022eqq,Korodi:2022ohn}, we do not consider the variance of the parameters within the ensemble of the utilized blocks of $N_{\rm events}$ events. This variance can be considered as a separate ``observable'', thus in the subsequent part, instead of this variance we plot the above discussed uncertainties with the central values obtained from the EPOS3 calculations. We nevertheless note here that this variance amounts to $1-1.5\%$ for central collisions and $0.5-1\%$ for peripheral collisions, where a range is indicated as this variance is the smallest for $\lambda$, larger for $\alpha$, and the largest for the radii. The centrality dependence comes from the fact that to reach stable parameters, a higher $N_{\rm events}$ value has to be considered, which decreases the variance.

\section{Results and discussion}\label{s:results}

\begin{figure*}
    \includegraphics[width = \linewidth]{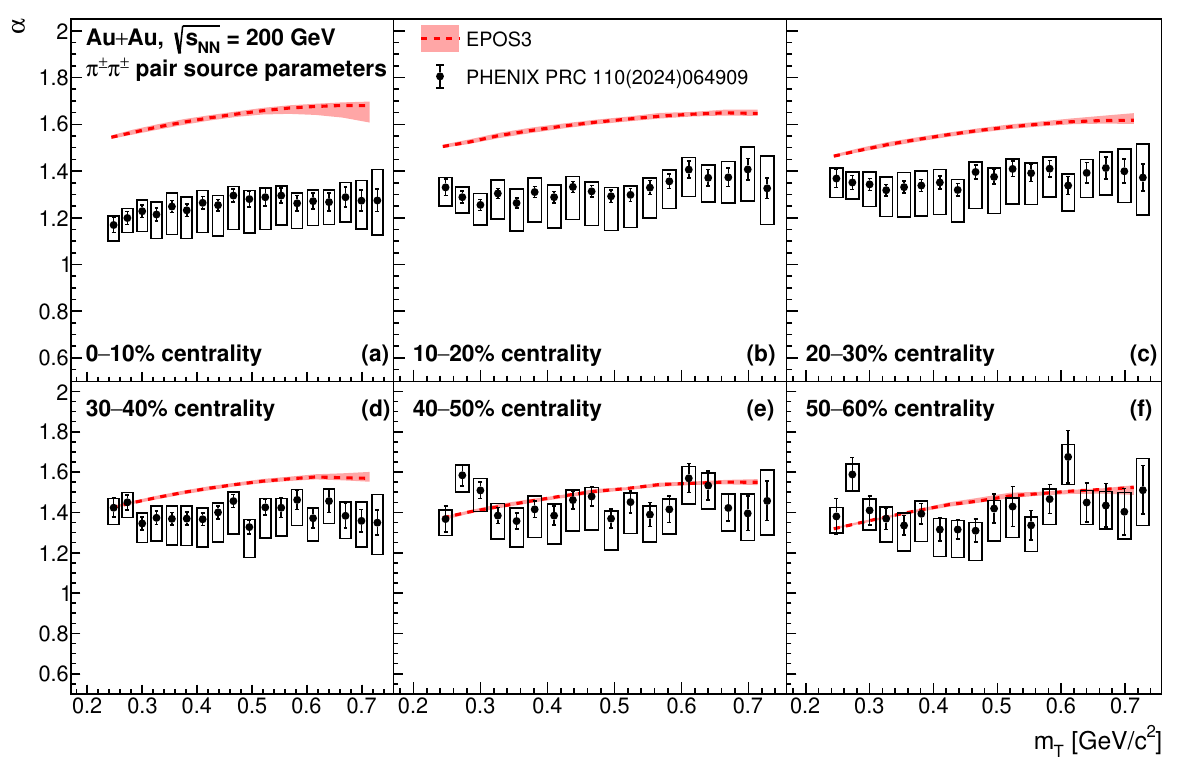}
    \caption{Lévy index $\alpha$ from EPOS3 (red line with band), compared to PHENIX data of Ref.~\cite{PHENIX:2024vjp} (black points), in six centrality ranges, all as a function of $m_T$. The red band shows the systematic uncertainties of the EPOS3 calculations. The statistical and systematic uncertainties of the PHENIX data points are shown as error bars and boxes, respectively.}\label{f:alpha}
\end{figure*}

\begin{figure}
    \includegraphics[width = \linewidth]{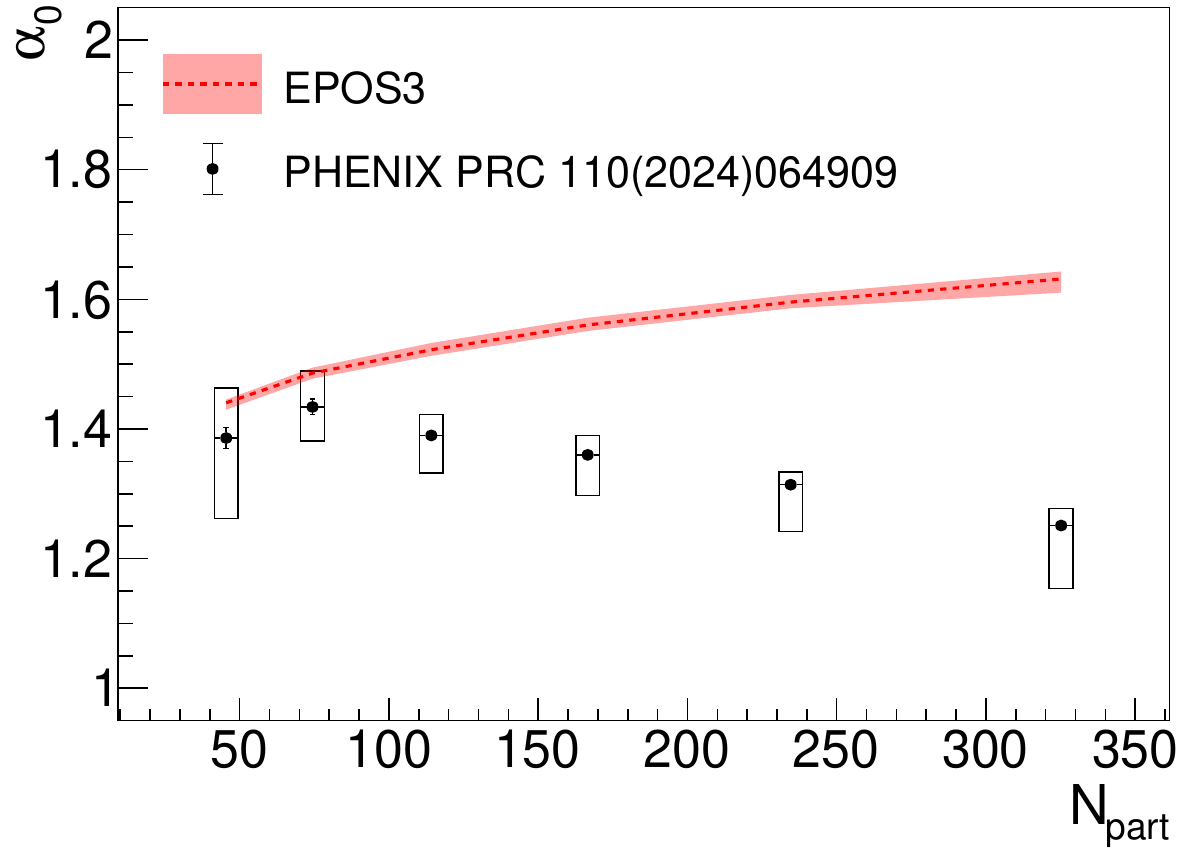}
    \caption{Lévy index $\alpha$ from EPOS3 (red line with band), compared to PHENIX data of Ref.~\cite{PHENIX:2024vjp} (black points), as a function of the number of participants $N_{\rm part}$.}\label{f:alpha0}
\end{figure}

Figure~\ref{f:alpha} shows the Lévy index $\alpha$ as a function of $m_T$, compared to PHENIX data of Ref.~\cite{PHENIX:2024vjp}. These plots indicate that while the trend of the data versus $m_T$ are well captured by the simulations, the change with centrality is not, and hence there is an increasing discrepancy between the absolute magnitude of $\alpha$ towards central collisions. In addition, Fig.~\ref{f:alpha0} shows the $m_T$-averaged ${\alpha_0=\left<\alpha(m_T)\right>}$ values versus the number of participants ($N_{\rm part}$)---the increasing discrepancy for central collisions is further illustrated by this figure. One effect that is not incorporated in EPOS3 and may be responsible for this effect is the scattering of charged pions induced by the Coulomb-interaction, as also mentioned in Ref.~\cite{Kincses:2024lnv}. This has been shown to affect the strength of quantum-statistical correlations in Ref.~\cite{Mohammed:2024pmg}, and one may expect a similar effect on the power-law tail of spatial distributions. Another possibility is the in-medium mass (and width) modifications, as also discussed in Ref.~\cite{PHENIX:2024vjp}---this may affect not just the correlation strength but also the spatial pair source as well. However, it is beyond the scope of present manuscript to investigate these effects in more detail.

\begin{figure*}
    \includegraphics[width = \linewidth]{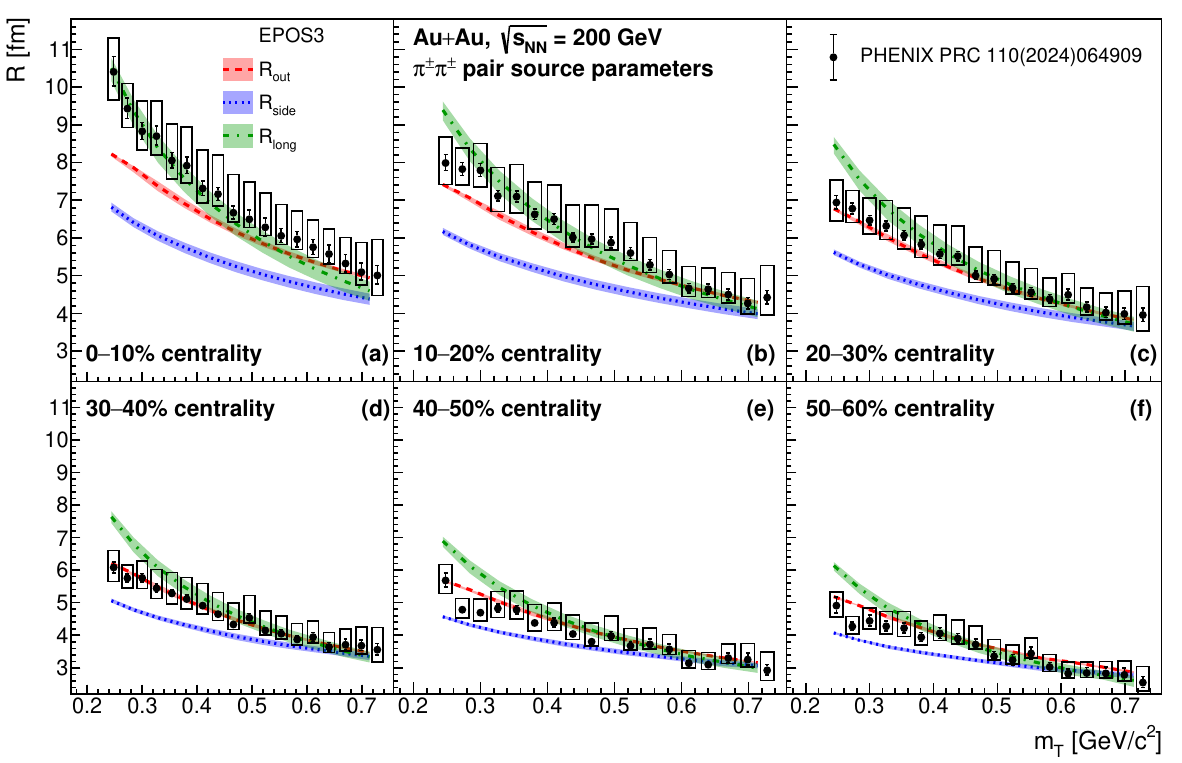}    
    \caption{Lévy scale parameters $R_{\rm out,side,long}$ from EPOS3  (red, blue, and green lines with bands, respectively), compared to PHENIX data of Ref.~\cite{PHENIX:2024vjp} (black points), in six centrality ranges, all as a function of $m_T$. The red, blue, and green bands show the systematic uncertainties of the EPOS3 calculations. The statistical and systematic uncertainties of the PHENIX data points are shown as error bars and boxes, respectively.}\label{f:R3D}
\end{figure*}

\begin{figure*}
    \includegraphics[width = \linewidth]{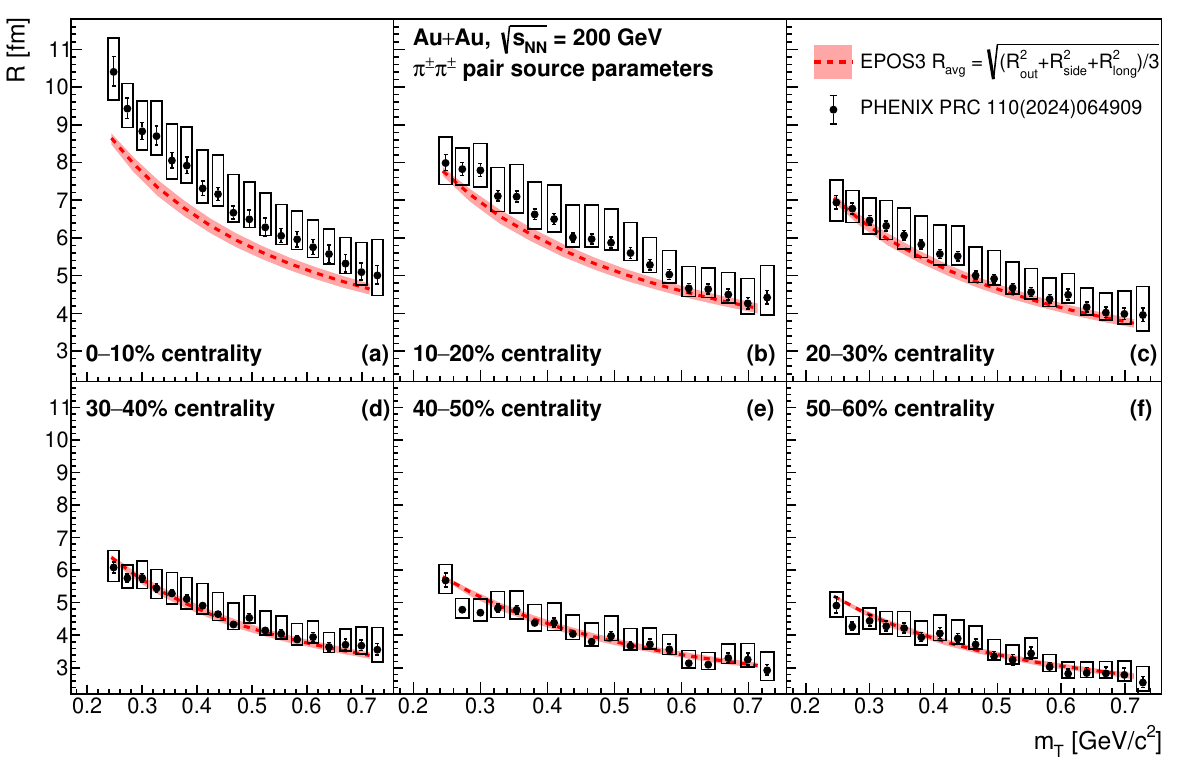}    
    \caption{Averaged Lévy scale parameter $R_{\rm avg}$ from EPOS3 (red line with band), compared to PHENIX data of Ref.~\cite{PHENIX:2024vjp} (black points), in six centrality ranges, all as a function of $m_T$. The red band shows the systematic uncertainties of the EPOS3 calculations. The statistical and systematic uncertainties of the PHENIX data points are shown as error bars and boxes, respectively.}\label{f:Ravg}
\end{figure*}

Figure~\ref{f:R3D} shows the Lévy scale parameters $R_{\rm out,side,long}$ as a function of $m_T$, compared to PHENIX data of Ref.~\cite{PHENIX:2024vjp}. The simulations capture both the trend and the absolute magnitude of the data as well, although the one-dimensional radii from the experiment tend to be closer to the largest of the three radii. To investigate this in more detail, Figure~\ref{f:Ravg} shows the averaged Lévy scale parameter ${R_{\rm avg}=\sqrt{(R_{\rm out}^2{+}R_{\rm side}^2{+}R_{\rm long}^2)/3}}$. It is apparent that this average scale is well described by EPOS3 for centralities above 20\%, and the discrepancy in the most central classes may be related to the same physics as what was discussed above for the $\alpha$ parameter.

\begin{figure*}
    \includegraphics[width = \linewidth]{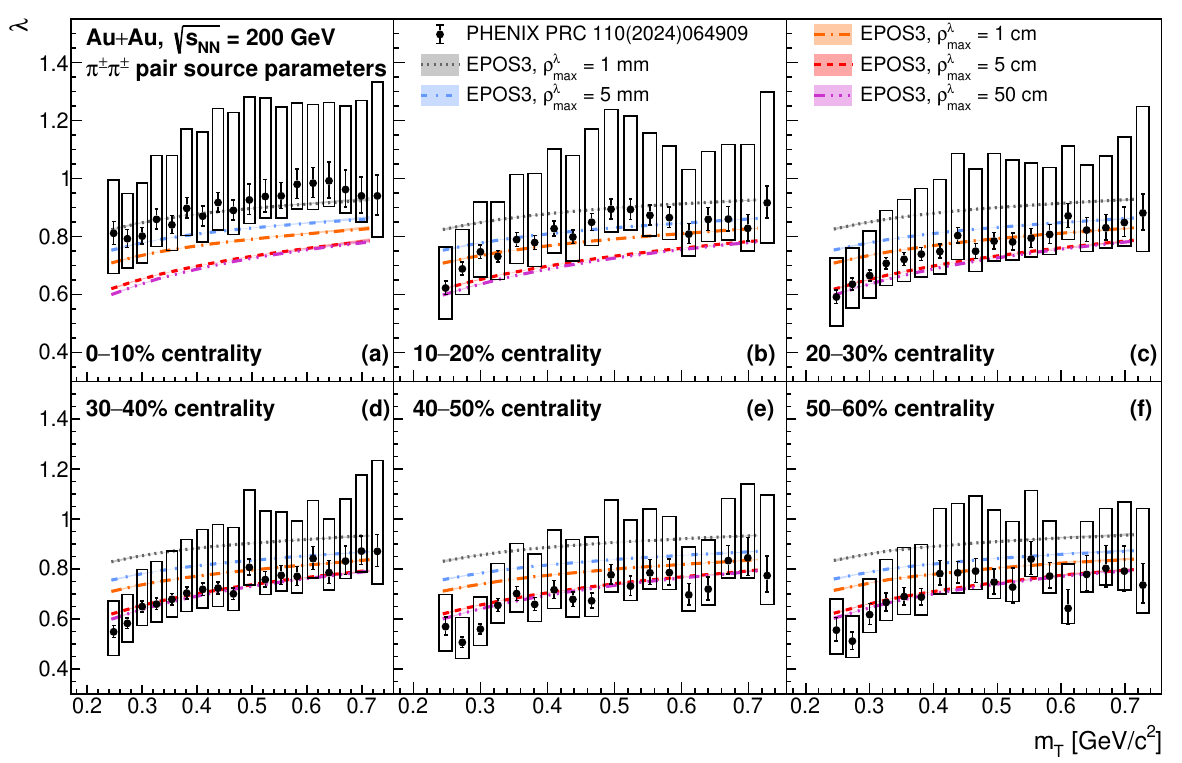}    
    \caption{Correlation strength parameters $\lambda$ from EPOS3 for various choices of $\rho^\lambda_{\rm max}$ (lines of different colors with bands), compared to PHENIX data of Ref.~\cite{PHENIX:2024vjp} (black points), in six centrality ranges, all as a function of $m_T$. The bands show the systematic uncertainties of the EPOS3 calculations. The statistical and systematic uncertainties of the PHENIX data points are shown as error bars and boxes, respectively.}\label{f:lambda}
\end{figure*}

Figure~\ref{f:lambda} shows the correlation strength parameters $\lambda$ as a function of $m_T$, for various choices of $\rho^\lambda_{\rm max}$, compared to PHENIX data of Ref.~\cite{PHENIX:2024vjp}. These plots indicate that, similarly to the case of $\alpha$ and $R$, EPOS3 captures the trend of the data, and describes their magnitude in peripheral collisions, while an increasing discrepancy appears either for central collisions, or for peripheral collisions, depending on the chosen normalization range: $\rho^\lambda_{\rm max}=$ 1 mm works for 0--10\% centrality, 5 mm for 10--20\% centrality, while 5 or 50 cm works for the more peripheral cases. This may be caused either by an incomplete handling of hadrons and their dynamics in EPOS3, in particular several neglected resonances and decay channels that can affect the source shape and strength, or the lack of Coulomb-scattering, as discussed above. Alternatively, it is also possible that different experimental pair resolutions have to be considered for different centralities. (Note that $\rho^\lambda_{\rm max}$ only affects $\lambda$, not the other source parameters.) However, the fact that the $m_T$-dependence is well captured by EPOS3 is interesting on its own.

\begin{figure*}
    \includegraphics[width = \linewidth]{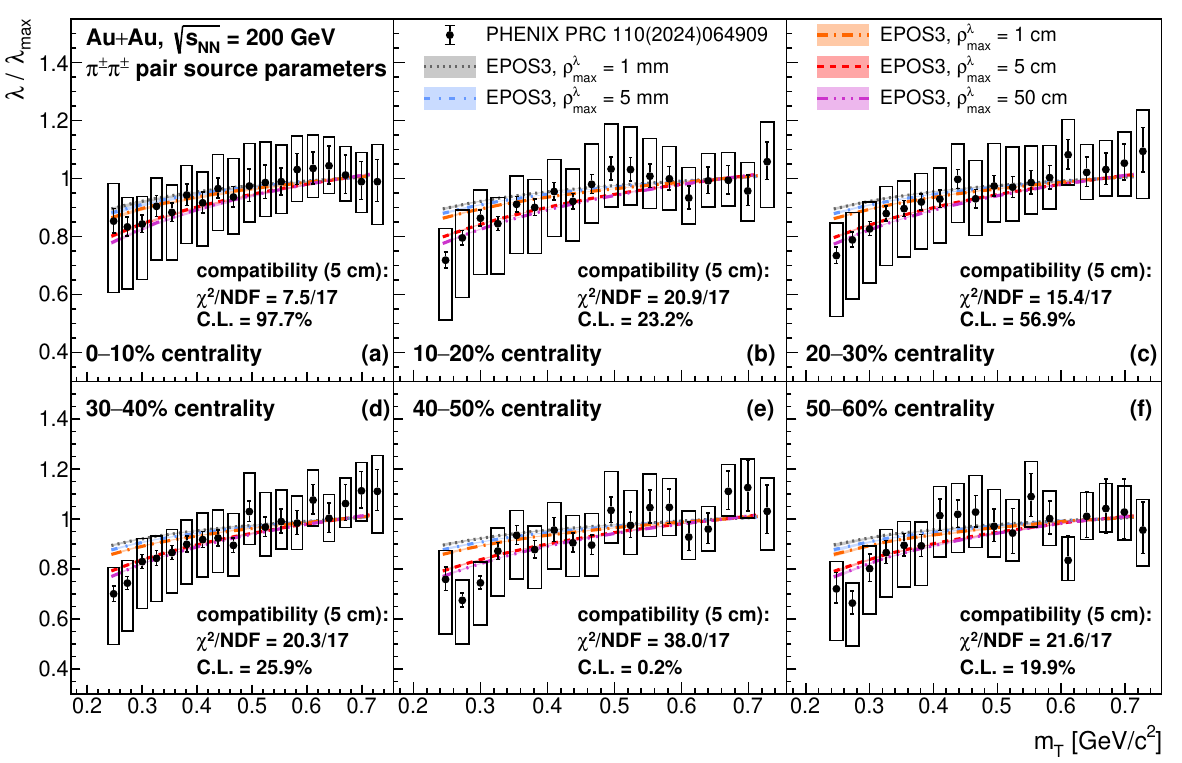}   
    \caption{Scaled correlation strength parameters $\lambda/\lambda_{\rm max}$ from EPOS3 for various choices of $\rho^\lambda_{\rm max}$ (lines of different colors with bands), compared to PHENIX data of Ref.~\cite{PHENIX:2024vjp} (black points), in six centrality ranges, all as a function of $m_T$. The bands show the systematical uncertainties of the EPOS3 calculations. The statistical and systematical uncertainties of the PHENIX data points are shown as error bars and boxes, respectively. The compatibility of the data and the EPOS3 results for the $\rho^\lambda_{\rm max}=5$ cm choice is also shown for each centrality, assessed through the $\chi^2$, number of degrees of freedom (NDF, number of data points in this case) and confidence level (C.L.) values, based on statistical uncertainties plus 5\% of the systematical uncertainties of the data points.}\label{f:lambdamax}
\end{figure*}

Similarly to Ref.~\cite{PHENIX:2024vjp}, we also investigated the scaled correlation strength parameter $\lambda/\lambda_{\rm max}$, as this removes the ambiguity of overall resonance fraction calculations, and focuses on the kinematical variation of the core fraction. Figure~\ref{f:lambdamax} shows $\lambda/\lambda_{\rm max}$ as a function of $m_T$, for various choices of $\rho^\lambda_{\rm max}$, compared to PHENIX data of Ref.~\cite{PHENIX:2024vjp}. The EPOS3 results appear to be compatible with the data for most of the $\rho^\lambda_{\rm max}$ choices. This compatibility was in particular assessed for the ${\rho^\lambda_{\rm max}= 5\textnormal{ cm}}$ case through the $\chi^2$ (calculated as the squared sum of their difference, scaled by the data uncertainty) and the corresponding confidence level values (abbreviated as C.L.; alternatively, these can also be called $p$-values). For all centralities, compatibility can be established (${\rm{C.L.} > 0.1\%}$). The data uncertainties were taken as the statistical uncertainties plus 5\% of the systematic uncertainties of the data points, which is a rather conservative estimate of the fluctuating part of the systematic uncertainties. This confirms quantitatively that the EPOS3 results provide a statistically acceptable description of the PHENIX $\lambda/\lambda_{\rm max}$ data of Ref.~\cite{PHENIX:2024vjp}, without the need for any modification in the default settings. Note however, that EPOS may slightly underestimate radial flow, indicated by Fig. 58. of Ref.~\cite{Werner:2023mod}. This may be an important detail to be investigated in the future, as the shape of $\lambda/\lambda_{\rm max}$ versus $m_T$ is strongly affected by radial flow~\cite{Vance:1998wd}.

\section{Summary and outlook}\label{s:summary}
We presented a three-dimensional femtoscopic analysis of pion emission sources in Au+Au collisions using the EPOS3 event generator, and compared the extracted Lévy source parameters to recent PHENIX measurements. The study focused on the transverse-mass and centrality dependence of the Lévy index, the Lévy scale parameters, and the correlation strength. The EPOS3 calculations result in pair sources compatible with the assumption of Lévy distributions, underlined by the fit quality and the small dependence on the fit range. The calculations furthermore reproduce the qualitative trends observed in data, including the overall decrease of source scales with increasing transverse mass and the expected centrality dependence, although quantitative deviations remain, particularly in the longitudinal direction. Furthermore, the EPOS3 results reproduce the scaled correlation strengths for all centralities. On the other hand, while agreement is found for peripheral and mid-central collisions in case of the Lévy index, there is an increasing discrepancy towards more central collisions.

One of the key messages from these comparisons is that unlike the interpretations given in Ref.~\cite{PHENIX:2024vjp}, the $\lambda/\lambda_{\rm max}$ data can be explained by EPOS3 without the need for in-medium modifications. On the other hand, unscaled $\lambda$ data cannot be explained, and possible reasons for this include experimental spatial resolution (reflected in the $\rho^\lambda_{\rm max}$ choice when comparing calculations to data), Coulomb scattering or hadronic dynamics. The most important and clearest discrepancy between calculations and data appears for the $\alpha$ parameter, in the case of central collisions. Subsequent investigations may reveal that Coulomb scattering or in-medium modifications of hadronic properties may be utilized to explain these differences.

Thus, our results provide new insight into the spatial structure of the pion-emitting source at freeze-out and demonstrate that hydrodynamics-based transport models coupled with hadronic dynamics such as EPOS3 can capture key femtoscopic observables within their current parameterizations, while also pointing towards the need of incorporating additional effects. Future work will focus on identifying these effects, responsible for the differences between data and simulation.


\bmhead{Acknowledgements}

This research was funded by the NKFIH grants TKP2021-NKTA-64, PD-146589, K-146913, K-138136, and NKKP ADVANCED 152097. The authors thank T. Csörgő and G. Kasza for the insightful discussions during the 2025 Days of Femtoscopy conference. 


\bibliography{sn-bibliography}

\end{document}